\documentclass{elsart}
\usepackage{natbib}
\usepackage{amsmath,amsfonts,amssymb,color,graphicx}

\newcommand{\La}{\Lambda}

\newcommand{\f}{\frac}

\def\be{\begin{equation}}
\def\ee{\end{equation}}
\begin{document}

\begin{frontmatter}

\title{Strange Star Equation of State With a Modified
Richardson Potential.}
\author[presi,dst]{Manjari Bagchi},
\ead{mnj2003@vsnl.net}
\author[iucaa]{Subharthi Ray},
\author[presi,dst]{Mira Dey},
\author[presi,dst]{Jishnu Dey}
\address[presi]{Presidency College, 86/1 College Street, Kolkata 700073, India}
\address[iucaa]{IUCAA, Post bag 4, Ganeshkhind, Pune 411007, India}
\thanks[dst]{Work supported in part by DST grant no.
SP/S2/K-03/2001, Govt. of India.}

\begin{abstract}
Richardson potential is an phenomenological interquark
interaction taking care of two aspects of QCD, namely the
asymptotic freedom and the confinement. The original potential
has a scale parameter having value $\sim400~MeV$ and is well
tested in hadronic property calculations. This potential was then
used in strange star calculation. Strange stars are very compact
stars composed  of strange quark matter $i.e.$ a very high density
strange quark phase consisting of deconfined u, d and s quarks.
Here the value of the scale parameter was taken as $100~MeV$. The
argument was that for a deconfined quark system like a strange
star, the scale parameter may have a value quite different from
that used in hadronic sector. To remove this discrepancy we
introduced two scale parameters in the potential, one for the
asymptotic freedom part and the other for the confining part.
With suitable values of the parameters, this modified potential
has been successfully used in both baryonic property and strange
star calculations. The Equation of States obtained with the
modified potential are also used to obtain mass-radius relations
for the strange stars.

\end{abstract}


\end{frontmatter}

\section{Introduction}
Two decades have passed since the existence of strange quark
matter (SQM) and strange stars (SS) were proposed by
\cite{witprd}, and even today it is still difficult to prove or
disprove the existence of SS with certainty. But some recent
observational signatures in X-ray and $\gamma$-ray astronomy
support the existence of SS \citep{okzhit,bignami,skin} inspiring
us to produce some new equation of states (EOS) for SQM. Present
work is an extension and improvement of the Realistic Strange
Star Model \cite{D98}.

\section{Realistic Strange Star (ReSS) Model and its modification}

Using the inverse of number of colors as an expansion parameter
in QCD \citep{thooft}, baryon like systems as SS can be explored
in relativistic tree level calculation with an average interquark
potential. Richardson potential is a phenomenological one used
for this purpose. The original potential has the asymptotic
freedom (AF) and confinement built in \citep{rich} and is used
successfully in mesonic \citep{cv} and baryonic \citep{ddt}
property calculations. The potential is : \be V_{ij} = \f{12
\pi}{27}\f{1}{ln(1 +  {({\bf k}_i- {\bf k}_j)}^2 /\La
^2)}\f{1}{{({\bf k}_i - {\bf k}_j)}^2} \label{eq:V} \ee $({\bf
k}_i - {\bf k}_j)$ being the momentum transfer. The scale
parameter ${\La}$ is  $\sim400~MeV$ for hadron phenomenology. But
in SS calculation ${\La}$ is taken $\sim100~MeV$. This
discrepancy ultimately leads us to modify the potential. The
potential is screened due to gluon propagation in a medium and
$({\bf k}_i - {\bf k}_j)^2$ is replaced by $[{({\bf k}_i - {\bf
k}_j)^2}+D^{-2}]$. Inverse screening length $D^{-1}$ ($i.e.$ gluon
mass $m_g$) is \citep{D98}:
\begin{equation} (D^{-1})^2 ~=~ \frac{2 \alpha_0}{\pi}
\sum_{i=u,d,s,}k^f_i \sqrt{(k^f_i)^2 + M_i^2} \label{eq:gm}
\end{equation} The Fermi momentum ($k^f_i$) is
related to the corresponding number density:
\begin{equation}
{k^f_i} = (n _i \pi^2)^{1/3} \label{eq:kf}
\end{equation}
Perturbative quark gluon coupling, $\alpha_0$ has value $\sim0.2$.
The quark mass, $M_i$ is taken to be density dependent to restore
chiral symmetry at high density :
 \be M_i = m_i + M_Q ~~sech\left( \f{n_B}{N n _0}\right),
\;\;~~~ i = u, d, s. \label{eq:qm} \ee where the constituent
quark mass $M_Q$ lies between 300-350 $MeV$ for each quark, $n_B
= (n _u+n _d+n _s)/3$ is the baryon number density, $n _0$ is the
normal nuclear matter density and $N$ is a parameter. The current
quark masses, $m_i$ are taken as 4, 7 and 150 $MeV$ for u, d, s
quarks respectively. From the charge neutrality and beta
equlibrium conditions, $k_i^f$ and $\mu_i$ (chemical potentials)
of quarks and electron are obtained as a function of $n_B$ and
performing a relativistic Hartree-Fock calculation, the energy
density $\epsilon$ of the SQM is obtained. The first law of
thermodynamics gives the EOS of SQM at zero temperature as:  \be
P~=~\sum_i({\mu_i}{n_i}-{\epsilon_i}) \label{eq:eos} \ee To obtain
a realistic EOS, the model parameters $\alpha_0$, $M_Q$ and $N$
are chosen in such a way that the minimum value of energy per
baryon ($E/A~\equiv~{\epsilon}/{n_B}$) for uds quark matter is
less than that of the most stable element $Fe^{56}$ $i.e.$ 930
$MeV$. Thus uds quark matter can construct stable stars. The
minimum value of $E/A$ is obtained at the star surface where the
pressure is zero. The presence of zero pressure indicates the
existence of a sharp surface of the strange star in contrast to
fuzzy surfaces of the neutron stars. The surface is sharp since
strong interaction dictates the deconfinement point. On the
contrary; for the same parameters, the minimum value of $E/A$ for
ud quark matter is greater than that of $Fe^{56}$ so that
$Fe^{56}$ remains the most stable element in the non-strange
world. The EOS for strange quark matter is now used to get the
Mass-Radius configuration of the strange star by solving the TOV
equations with appropriate boundary conditions.

In the present work, we have assumed that the scales for the two
phenomena AF and confinement are different and the modified
Richardson potential is:
\begin{eqnarray}
\nonumber V_{ij} &=& \frac{12 \pi}{27}\left[\frac{1}{({\bf k}_i -
{\bf k}_j)^2{\rm ln}(1 + \frac{{({\bf k}_i- {\bf k}_j)}^2}{\Lambda
^2})}-\frac{\Lambda^2}{({\bf k}_i- {\bf k}_j)^4}
+\frac{{\Lambda^\prime}^2}{({{\bf k}_i- {\bf k}_j})^4}\right]
\label{eq:Vtl}
\end{eqnarray}

Here ${\Lambda}$ is the scale representing the AF as the first two
terms are asymptotically zero for large $({{\bf k}_i- {\bf
k}_j})$. ${\Lambda}^{ \prime}$ is the scale corresponding to
confinement as the third term reduces to a linear confinement for
small $({\bf k}_i - {\bf k}_j)$. From baryonic property
calculations \citep{mmsj}, we have found the appropriate values of
the parameters as ${\Lambda}~=~100 ~MeV$ and
${\Lambda}^{\prime}~=~350 ~MeV$. So we have applied the modified
potential with these values of the parameters in SS calculations.

\section{Results}

\begin{table}[htbp]
\begin{center}
\caption{Different EOSs using the modified potential. ${\Lambda}$
is always 100 $MeV$. ${\epsilon}_c/c^2$ is the central density.
The present value of $\alpha_0$ is 0.55-0.65 as compared to 0.20
of \citet{D98}. We have to chose a larger value of $\alpha_0$ as
the confinement parameter has larger value. Such a large value of
$\alpha_0$ is in well agreement with theoretical results
\citep{denalph}. } \vskip 0.5 cm \label{tab:allmrnew}
\begin{tabular}{|c|c|c|c|c|c|c|c|c|c|}
\hline
EOS&${\Lambda}^{\prime}$&$M_q$&$N$&${\alpha_0}$&${(\frac{E}{A})}^{uds}_{min}$
&${(\frac{E}{A})}^{ud}_{min}$ &$M_{max}$& R & ${\epsilon}_c/c^2$\\
  &  &  & &&  & &  &  for  & for $M_{max}$ \\
&  &  & &&  & &  &  $M_{max}$ &  \\
 &  &  & &&  & &  &  & $10^{14}$\\
&$MeV$&$MeV$ & & &$MeV$& $MeV$&$M_\odot$& $km$& $gm~cm^{-3}$\\
\hline \hline

A& 350&325 &3.0 &.55 &874 &942& 1.53 &7.41 & 42.98 \\
\hline

B& 350&325 &3.0 &.65  &907&942&1.47&7.14 &  45.48 \\ \hline

C& 300&325 &2.7 &.55 &877 &942 &1.55 &7.53& 41.46  \\ \hline

D & 300&325 &3.0 &.55 &906&950 &1.46 &7.11&  46.14\\ \hline

E &300&335 &3.0 &.55 &912 &973 &1.45&7.02 & 47.65  \\ \hline

F &350&345 &3.0 &.65 &920& 987& 1.44& 6.97 & 48.07 \\
\hline \hline
\end{tabular}
\end{center}
\end{table}

We have obtained different EOSs for different set of parameters
(see table \ref{tab:allmrnew}). Though baryonic property
calculation tells that ${\Lambda}^{\prime}$ should be $350~MeV$,
we have determined EOSs (C, D, E) also for
${\Lambda}^{\prime}~=~300~MeV$ only for theoretical interest
$i.e.$ to see the effect of changing other parameters of the
model. The other EOSs (A, B, F) are indeed for
${\Lambda}^{\prime}~=~350~MeV$. Comparing with \citet{D98}, it is
clear that the nature of EOSs and M - R relations do not change
appreciably by the use of the modified potential if we adjust
other model parameters suitably; but the value of $m_g$ is
significantly larger as we have to increase the value of
$\alpha_0$ to counter balance the increase in
${\Lambda}^{\prime}$. The compactness (M/R ratio) of the stars
have increased for the new EOSs than that obtained by \citet{D98}.
Zero pressure is obtained for $n_B>4~n_0$ for all EOSs implying
that in a strange star, $n_B$ can not be less than $4~n_0$. The
central density may have different values giving rise to stars of
different masses, but for standard masses $i.e.$ around 1.4-1.5
$M_{\odot}$, it becomes $\sim~13~n_0$. This high density
indicates a high pressure.

\end{document}